%% file: paper.tex
\documentclass[conference]{IEEEtran}

\usepackage[normalem]{ulem}
\usepackage{adjustbox}
\usepackage{multirow}
\usepackage{pbox}
\usepackage{xspace}
\usepackage{url}
\usepackage{cite}
\usepackage{amsmath,amssymb,amsfonts, mathtools}
\usepackage{algorithmic}
\usepackage{subcaption}
\usepackage[linesnumbered,ruled,vlined]{algorithm2e}
\usepackage{graphicx}
\usepackage{textcomp}
\usepackage{xcolor}
\usepackage{comment}
\def\BibTeX{{\rm B\kern-.05em{\sc i\kern-.025em b}\kern-.08em
    T\kern-.1667em\lower.7ex\hbox{E}\kern-.125emX}}

\newcommand{\nvm}{PMEM\xspace}
\newcommand{\sys}{Cori\xspace}

\makeatletter
\def\tightitemize{\ifnum \@itemdepth >3 \@toodeep\else \advance\@itemdepth \@ne
\edef\@itemitem{labelitem\romannumeral\the\@itemdepth}%
\list{\csname\@itemitem\endcsname}{\setlength{\topsep}{-\parskip}\setlength{\parsep}{0in}\setlength{\itemsep}{0in}\setlength{\parskip}{0in}\def\makelabel##1{\hss\llap{##1}}}\fi}

\def\tightenumerate{%
\ifnum \@enumdepth >\thr@@\@toodeep\else
\advance\@enumdepth\@ne
\edef\@enumctr{enum\romannumeral\the\@enumdepth}%
\expandafter
\list
\csname label\@enumctr\endcsname
{\setlength{\topsep}{-\parskip}\setlength{\parsep}{0in}\setlength{\itemsep}{0in}\setlength{\parskip}{0in}\usecounter\@enumctr\def\makelabel##1{\hss\llap{##1}}}%
\fi}

\makeatother

\newenvironment{tightbulletitemize}%
         {\begin{list}{$\bullet$}{%
                \setlength{\leftmargin}{10pt}
        \setlength{\itemsep}{0pt}%
        \setlength{\parsep}{0pt}%
        \setlength{\topsep}{0pt}%
        \setlength{\parskip}{0pt}%
        }%
 }%
{\end{list}}

\newcounter{tecounter}
\newenvironment{tightnumberenumerate}%
 {\begin{list}{\arabic{tecounter}.}{%
                \usecounter{tecounter}
                \setlength{\leftmargin}{10pt}
        \setlength{\itemsep}{0pt}%
        \setlength{\parsep}{0pt}%
        \setlength{\topsep}{0pt}%
        \setlength{\parskip}{0pt}%
        }%
 }%
{\end{list}}%

\begin{document}

\title{Tuning the Frequency of Periodic Data Movements over Hybrid Memory Systems}

\author{\IEEEauthorblockN{Thaleia Dimitra Doudali}
\IEEEauthorblockA{\textit{Georgia Institute of Technology} \\
thdoudali@gatech.edu}
\and
\IEEEauthorblockN{Daniel Zahka}
\IEEEauthorblockA{\textit{Georgia Institute of Technology} \\
dzahka3@gatech.edu}
\and
\IEEEauthorblockN{Ada Gavrilovska}
\IEEEauthorblockA{\textit{Georgia Institute of Technology} \\
ada@cc.gatech.edu}}

\maketitle
\thispagestyle{empty}

\input{abstract}


\input{introduction}

\input{methodology}

\input{motivation}
\input{solution}

\input{evaluation}

\input{related}
\input{summary}

{\bibliographystyle{abbrv} \bibliography{paper}}

\end{document}

%% file: abstract.tex
\begin{abstract}
  Emerging hybrid memory systems that comprise technologies such as Intel's Optane DC Persistent Memory, exhibit disparities in the access speeds and capacity ratios of their heterogeneous memory components. This breaks many assumptions and heuristics designed for traditional DRAM-only platforms. High application performance is feasible via dynamic data movement across memory units, which maximizes the capacity use of DRAM while ensuring efficient use of the aggregate system resources. Newly proposed solutions use performance models and machine intelligence to optimize which and how much data to move dynamically; however, the decision of when to move this data is based on empirical selection of time intervals, or left to the applications.
  Our 
  experimental evaluation shows that
  failure to properly configure the data movement frequency can lead to 10\%-100\% slow down for a given data movement policy; yet, there is no established methodology on how to properly configure this value for a given workload, platform and policy. 
  We propose \sys, a system-level tuning solution that identifies and extracts the necessary application-level data reuse information, and guides the selection of data movement frequency to deliver gains in 
  application performance and system resource efficiency.
  Experimental evaluation shows that \sys
configures data movement frequencies that provide application performance within 3\% of the optimal one, and that it can achieve this up to  5$\times$ more quickly than random or brute-force approaches. 
  System-level validation of \sys on a platform with DRAM and Intel's Optane DC PMEM confirms its practicality and tuning efficiency.
\end{abstract}

%% file: introduction.tex
\section{Introduction}
\label{sec:intro}

\noindent{\bf The Era of Massive Hybrid Memory Systems.} 
Big data analytics, machine learning workloads and data intensive scientific simulations need massive main memory capacities to accelerate data retrieval times and overall application performance. To satisfy this demand, memory hierarchies have become more complex, incorporating emerging memory technologies and disaggregation techniques in order to offset the skyrocketing cost that DRAM-only systems would impose. For example, the Intel\textsuperscript{\textregistered} Optane\textsuperscript{\tiny TM} DC Persistent Memory (\nvm) platform introduces a high-density, non volatile memory technology at least 3$\times$ {\it slower} than DRAM \cite{optane-intel, optane-swanson}, but also 2$\times$-3$\times$ {\it cheaper} than DRAM \cite{memverge}. Server configurations with 6 TBs of PMEM pagkaged together with 375 GB of DRAM, such as the Optane platform,
can significantly boost application performance with proper dynamic data management \cite{optane-maya-memsys19, optane-memsys19}.

\noindent{\bf Cache vs. Flat Memory Organization.} There are two primary ways to organize hybrid memory hierarchies. One is a vertical organization, where one memory type acts as a cache for the other and is managed by hardware. The other is a horizontal organization, where all memories `lay flat' and are managed by software -- the operating system or applications themselves. These correspond to the {\it Memory} and {\it App-direct} modes in Intel's Optane DC PMEM platform,
and each mode introduces different trade-offs
with respect to system resource efficiency and application performance. For instance, recent work has shown that the cache organization improves performance of graph applications~\cite{optane-vldb}. In contrast,
the flat organization allows for lower energy cost and higher bandwidth use~\cite{optane-maya-memsys19, optane-memsys19}, and 
a number of hardware and software techniques have recently been proposed to further improve the associated management 
overheads~\cite{HMA, thermostat, kommareddy-memsys19, mempod, hetero-os, hetero-visor}. 

\noindent{\bf Prefetching vs. Data Tiering.} Prefetching solutions speculatively bring select data from slower to faster memory, such as \nvm to DRAM. Recent advances in prefetching include novel access pattern prediction methods that use machine learning \cite{LMA} and dynamic windowing together with majority voting techniques \cite{leap}. In contrast, data tiering solutions dynamically rearrange data allocations across {\it flat} hybrid memory, such that frequently accessed data resides in fast memory (e.g., DRAM), thus maximizing its use. Recent advances in data tiering incorporate machine intelligent methods into the data selection and movement process~\cite{kleio}. 

\noindent{\bf Data tiering solutions} often include a {\it page scheduler} that monitors data access behavior, and periodically migrates data across hybrid memory tiers. While a significant body of research focuses on optimizing the selection of {\it which} data to move, there is little insight towards {\it when} that data should be moved. Focusing on the latter, Table~\ref{tbl:related_work} summarizes the operational frequencies of related data tiering solutions, whose difference in time ranges four orders of magnitude. These values are {\it empirically} tuned to meet the performance requirements of the specific pool of applications evaluated for their respective systems.

\begin{table}
\centering
\begin{adjustbox}{width=\linewidth}
\begin{tabular}{|c|c|c|}\hline
{\bf Solution} & {\bf Period Duration} & {\bf Requests per Period}  \\\hline
Thermostat \cite{thermostat} & 10 sec &  100,000  \\\hline
Nimble \cite{nimble} & 5 sec & 50,000 \\\hline
Ingens \cite{ingens} & 2 sec &  20,000  \\\hline
HMA \cite{HMA} & 1 sec &  10,000  \\\hline
Hetero-OS \cite{hetero-os}, -Visor \cite{hetero-visor} & 0.1 sec  & 1,000 \\\hline
Kleio \cite{kleio} & 0.01 sec & 100  \\\hline
Unimem \cite{Unimem} & MPI phase & N/A \\\hline
\end{tabular}
\end{adjustbox}
\vspace{1.0ex}
\caption{Frequency of data monitoring and movement across existing solutions mapped to our simulation-based analogy.}
\vspace{-0.2in}
\label{tbl:related_work}
\end{table}

\noindent{\bf Empirical tuning} of page scheduling frequency can miss significant performance improvements by not testing certain frequency ranges in an effort to minimize tuning overhead. For example, a common approach \cite{HMA, kommareddy-memsys19} is to experiment with period durations that are an order of magnitude apart, e.g., 0.01 sec, 0.1 sec and 1 sec, so as to identify in only three trials which offers the highest DRAM hitrate while maintaining reasonable data movement overhead. On the other hand, exploring all frequency choices leads to impractical tuning overheads. In addition, the periodic solutions in Table~\ref{tbl:related_work} fix their operational frequency at the system-level, so that they do not have to repeat the empirical tuning for every application. However, this can potentially leave a significant amount of unexploited performance for applications with data access behaviors and sizes that the empirical tuning did not consider.
Another approach is to completely rely on the application to explicitly control data allocation and movement, via 
use of specialized pragmas or {\tt malloc}-like APIs.
Such modified applications then explicitly control how 
the underlying system-level solution 
maintains the necessary state to dynamically manage data tiering across hybrid memory \cite{xmem, memkind, Unimem, tahoe}. 

\noindent{\bf Problem Statement.} Impractical tuning overheads and lack of insight force existing data tiering solutions to rely on empirical tuning of their operational frequency, or on application-level modifications suitable for specific execution models and APIs. 
As a result, for general scenarios where modifying the applications is not appropriate, there can be significant levels of performance that existing data tiering solutions do not realize across applications, due to their empirically-tuned and fixed operational frequency.

\noindent{\bf Paper Contributions.} To address this, we propose {\bf \sys} --  a  system-level solution for tuning the operational periods in page schedulers, that maximizes the effectiveness of the schedulers in terms of application performance and platform efficiency, and achieves that with low tuning overheads. 

\sys operates in an application and runtime-agnostic manner, and relies on observation-based insights to guide the frequency tuning process to a small number of viable candidates.
We demonstrate that \sys is effective, 
{\it irrespective} of the data access behavior and page scheduling effectiveness, 
and can be practically 
integrated into the existing hybrid memory management software stack.

\noindent The specific contributions of this paper are the following:
\begin{tightbulletitemize}
\item We demonstrate that current data tiering solutions can experience 10\%-100\% performance loss due to sub-optimal choice of their operational frequencies 
  (Section \ref{sec:motiv-perf}).
\item We identify a relationship among observable application properties -- their data reuse -- and the favorable scheduling periods
  (Section \ref{sec:motiv-reuse}).
\item We describe the design 
  of
  {\bf \sys}\footnote{The name is inspired by the ancient Greek mythology, where Cori (short for Terpsichore) was the muse of dance and daughter of Mnemosyne, the goddess of memory.} 
  and its frequency tuning methodology, for a simulation-based prototype and in real system settings. 
  (Section~\ref{sec:solution}).
\item We evaluate \sys, demonstrating its ability to identify operational frequencies which realize performance improvements within only 3\%
  from the ideal frequency selection, on average, across applications and page scheduling variations. \sys achieves this with 5$\times$ fewer number of tuning trials,
  compared to insight-less 
  tuning approaches (Sections~\ref{sec:eval-benefit},~\ref{sec:eval-cost}).
\item We validate \sys's insights, effectiveness and practicality on a real hardware testbed with DRAM and Intel's Optane DC \nvm (Section \ref{sec:eval-validate}).
\end{tightbulletitemize}

%% file: methodology.tex
\section{Methodology}
\label{sec:method}

\begin{table}
\centering
\begin{adjustbox}{width=\linewidth}
\begin{tabular}{|c|c|c|c|}\hline
{\bf Application} & {\bf Abbrv.} & {\bf Suite} & {\bf Domain}  \\\hline
Back Propagation & {\tt backprop} & Rodinia & Machine Learning   \\\hline
Kmeans  & {\tt kmeans} & Rodinia & Machine Learning   \\\hline 
HotSpot & {\tt hotspot} & Rodinia & Physics Simulation \\\hline
LU Decomposition & {\tt lud} & Rodinia & Linear Algebra  \\\hline
Breadth-First Search  & {\tt bfs} & Rodinia & Graph Algorithms \\\hline
B+Tree  & {\tt bptree} & Rodinia & Databases \\\hline
Pennant  & {\tt pennant} & Coral-2 & Hydrodynamics \\\hline
Quicksilver & {\tt quicksilver} & Coral-2 & Monte-Carlo \\\hline
CP Decomposition & {\tt cpd} & ParTI! & Sparse Tensors \\\hline
\end{tabular}
\end{adjustbox}
\vspace{1.0ex}
\caption{Applications used in experiments.}
\label{tbl:apps}
\vspace{-2.0ex}
\end{table}

\noindent{\bf Applications.} Table \ref{tbl:apps} summarizes the applications that we selected for experimental evaluation from the Rodinia~\cite{rodinia}, Coral-2~\cite{coral2} and ParTI!~\cite{parti} benchmark suites. The selected benchmarks and mini-apps cover a wide range of application domains and memory access patterns.

\subsection{Optane DC PMEM Platform}
\label{sec:method-optane}
We have access to a server with Intel Optane DC Persistent Memory Modules (\nvm), which we configure in {\it App Direct} mode. The machine contains 375 GB of DRAM and 6 TB of \nvm.  
We implement a page migration module\footnote{https://github.com/GTkernel/x86-Linux-Page-Scheduler.git} for Linux kernel version 5.4 that attaches to a target process and periodically selects 4 KB pages to move between DRAM and \nvm. Every period, we identify page accesses using the available OS-level information, as also done in~\cite{hetero-visor, hetero-os}. The module determines which pages were accessed by scanning the target's page table entries and recording whether or not each accessed bit was set during that period.  To estimate the page hotness, we calculate the exponential moving average (with a certain smoothing factor) of the page's accessed bit history and compare it with a hotness threshold that classifies a page as hot or cold, as also done in \cite{ingens}. Then, utilizing the {\tt move\_pages()} function from the kernel's NUMA-based migration API, we asynchronously move hot pages to DRAM and cold pages to \nvm. The kernel module dynamically adjusts the page migration cutoff, dividing the process memory footprint across DRAM and PMEM at a certain capacity ratio.

\subsection{Simulation}
\label{sec:method-sim}

\noindent{\bf Memory Access Trace Collection.} We use Intel's Pin~\cite{intel-pin} dynamic binary instrumentation tool to capture the memory address of the last level cache misses out of a simulated three level data cache hierarchy. In order to allow for reasonable trace sizes and analysis times we simulate a cache hierarchy of smaller but proportional capacity ratio to the Intel Optane DC \nvm platform. Then we fix the application data inputs such that we observe similar last level cache miss rate to application execution in the native \nvm platform.

\noindent{\bf Hybrid Memory System.} 
We develop a Python-based simulation environment\footnote{https://github.com/GTkernel/cori-sim.git}
 that allows fast trace-based analysis similar to \cite{HMA, kleio}. In particular, we assume a flat organization of fast (e.g., DRAM) and slow (e.g. \nvm) memory, similar to the App Direct mode configuration of the Intel Optane platform. Following the observed \nvm access speeds \cite{optane-swanson} we set a 1:3 latency and 1:0.37 bandwidth ratio between the fast and slow memory. We assume that the overall capacity of the memory system is equal to an application's memory footprint. Since we are not using cycle-accurate simulation, we assume that a period is the time duration when a fixed number of memory requests are issued, e.g., 1,000 requests per period. To estimate the runtime we aggregate the access latency of the memory requests for their coresponding memory allocation across periods. In addition, we account for any limited bandwidth availability, by injecting appropriate delays given the number of memory requests serviced over a window of time. Finally, we add constant delays for every page migration and start of a period to account for the overhead of the page scheduler itself, using the proposed values in~\cite{HMA, kommareddy-memsys19}.

\noindent{\bf Page Scheduler.} We extend the Python-based simulation with a page scheduler that periodically aggregates per page access counts from the collected access trace and migrates pages between fast and slow memory. The initial page allocation is done in an interleaved manner across memories, which is typical for NUMA systems. Every period the page scheduler identifies the pages that are frequently accessed (hot) and moves to fast memory any hot pages that reside in slow memory, replacing any recently used (LRU) pages. The number of page migrations per period is capped by the available fast memory capacity, since hot and LRU pages are swapped across hybrid memory. These page swaps happen asynchronously, assuming DMA support, and sequentially in order of (hot,~LRU) page pairs.

We refer to this type of page scheduler, that makes a selection of page migrations using access history, as a {\bf reactive} page scheduler, since it `reacts' to the changes in the memory access pattern, as also done in \cite{thermostat, nimble, ingens, HMA, hetero-visor, hetero-os}. We also simulate a {\bf predictive} page scheduler, that predicts memory access patterns, thus makes a more sophisticated page migration selection or even has a-priori knowledge of the access pattern, as described as the oracular baseline in \cite{kleio, HMA}. This simulation configures the reactive page scheduler to act upon a single period of past access history, and similarly the predictive page scheduler to make an access pattern prediction for the upcoming period.

\noindent{\bf Comparison with existing solutions} aims to capture the application performance impact caused only by the selection of {\it when} to move data, not which and how much data to move. For this purpose we assume the aforementioned page scheduling implementations and evaluate upon the data movement frequencies of existing solutions, as summarized in Table~\ref{tbl:related_work}. Since these proposed values vary across orders of magnitude, we create corresponding period durations that map to our previously described runtime simulation.

%% file: motivation.tex
\begin{figure}
\centerline{\includegraphics[scale=0.55]{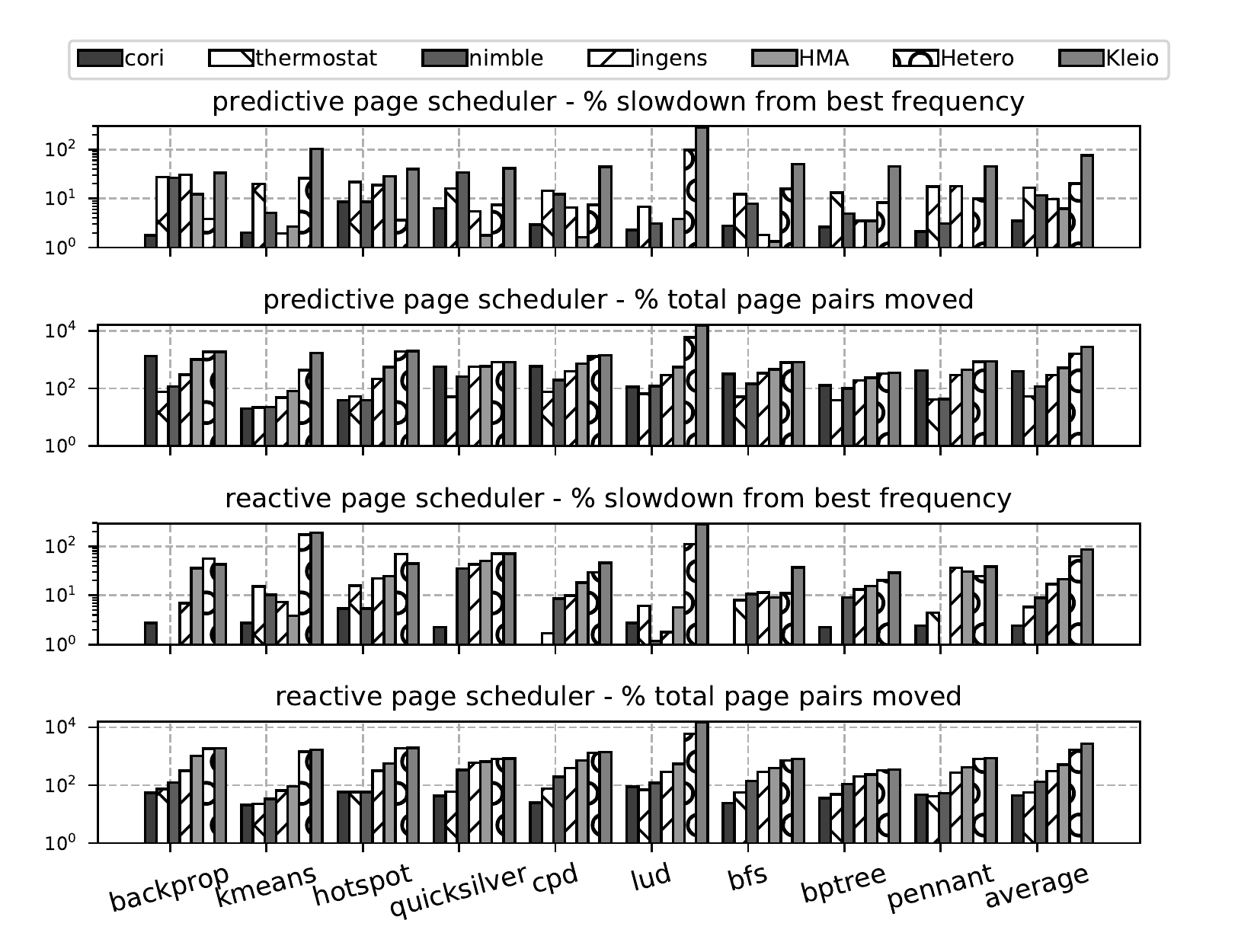}}
\vspace{-2.0ex}
\caption{Performance comparison of a predictive and reactive page scheduler across operational frequencies of existing solutions and the proposed solution \sys, given a simulated hybrid memory system with DRAM and \nvm at a 20\%:80\% capacity ratio.}
\vspace{-3.0ex}
\label{fig:perf_gap}
\end{figure}
\begin{figure*}[htbp]
\centering
\includegraphics[scale=0.65]{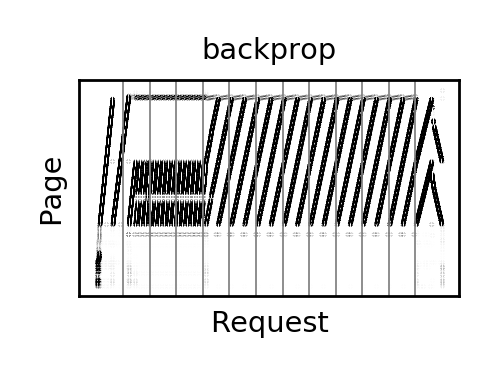}
\includegraphics[scale=0.65]{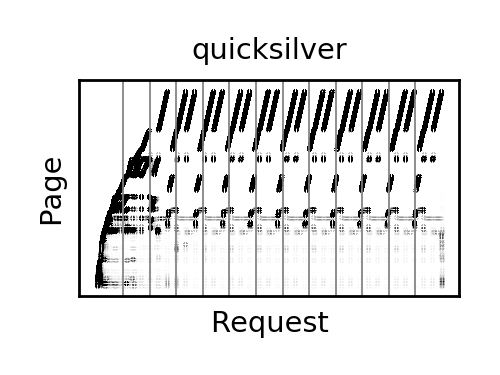}
\includegraphics[scale=0.65]{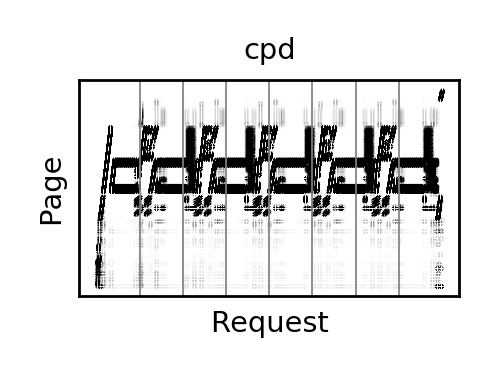}
\includegraphics[scale=0.65]{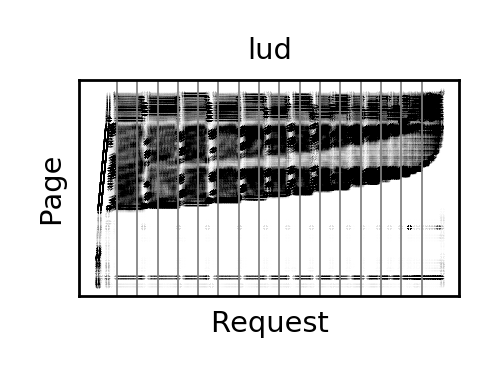}
\includegraphics[scale=0.65]{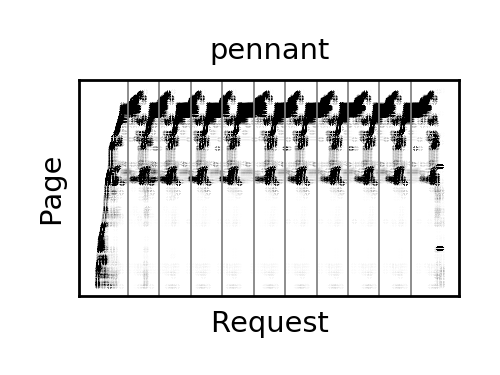}
\vspace{-2.0ex}
\caption{Representative memory access traces. The vertical lines correspond to the fixed period boundaries that provide best performance, as selected by \sys.}
\vspace{-2.0ex}
\label{fig:traces}
\end{figure*}

\begin{figure*}[htbp]
\centering
\includegraphics[scale=0.6]{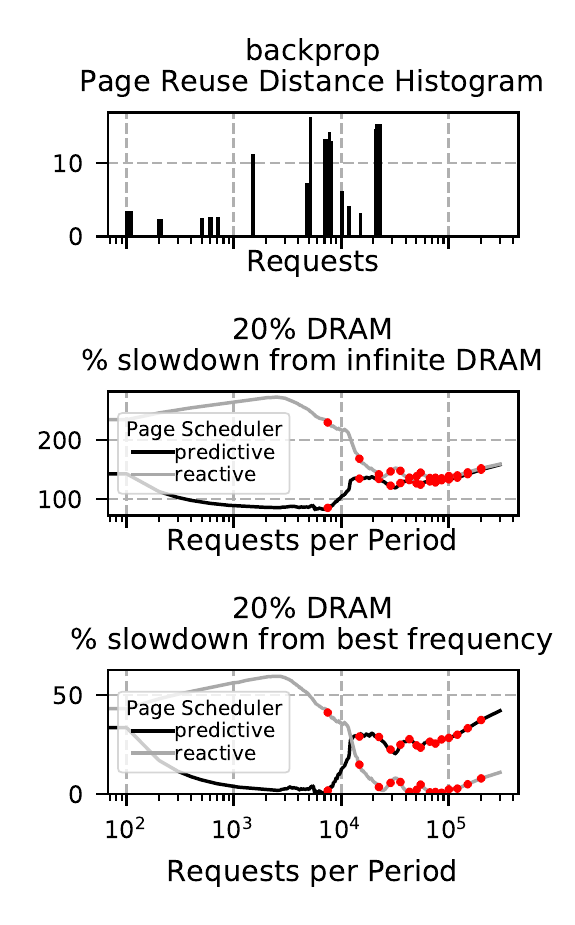}
\includegraphics[scale=0.6]{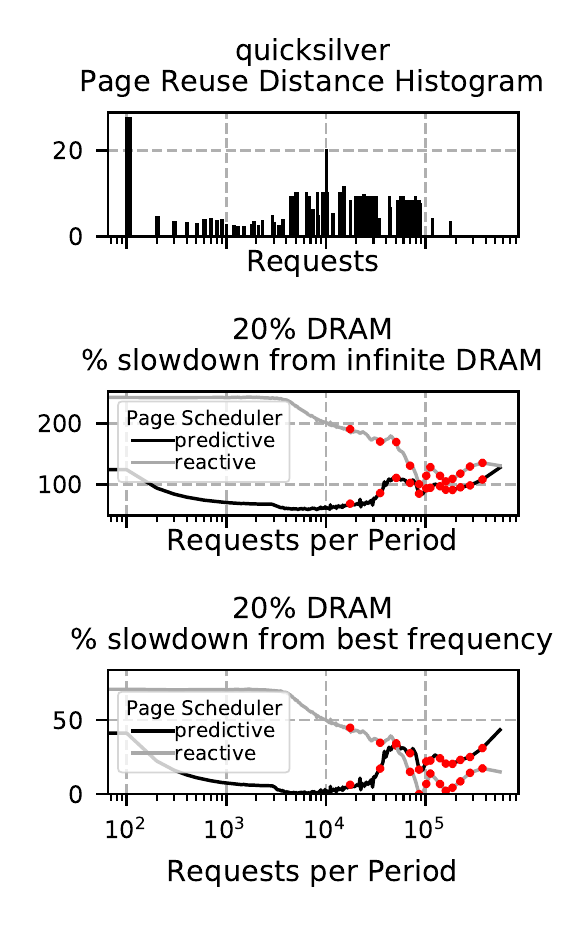}
\includegraphics[scale=0.6]{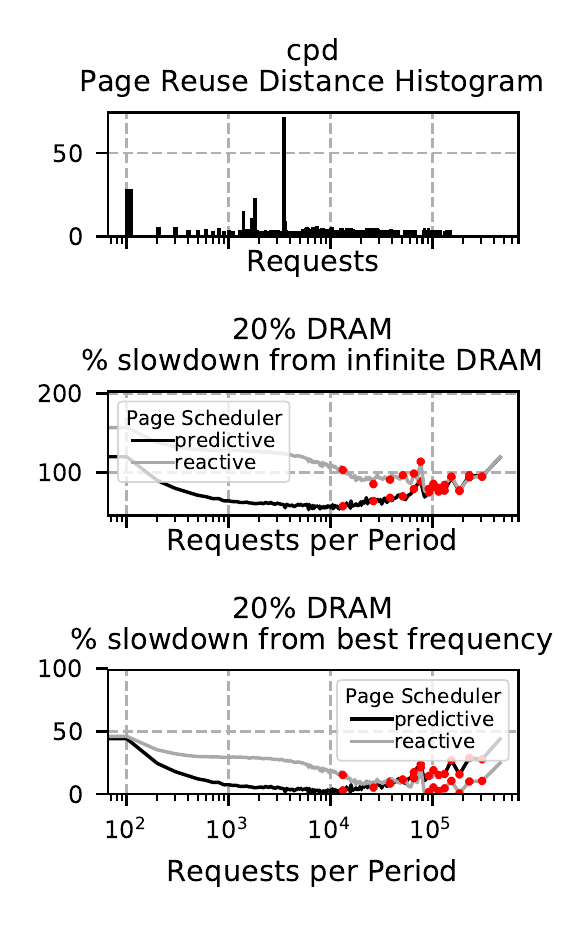}
\includegraphics[scale=0.6]{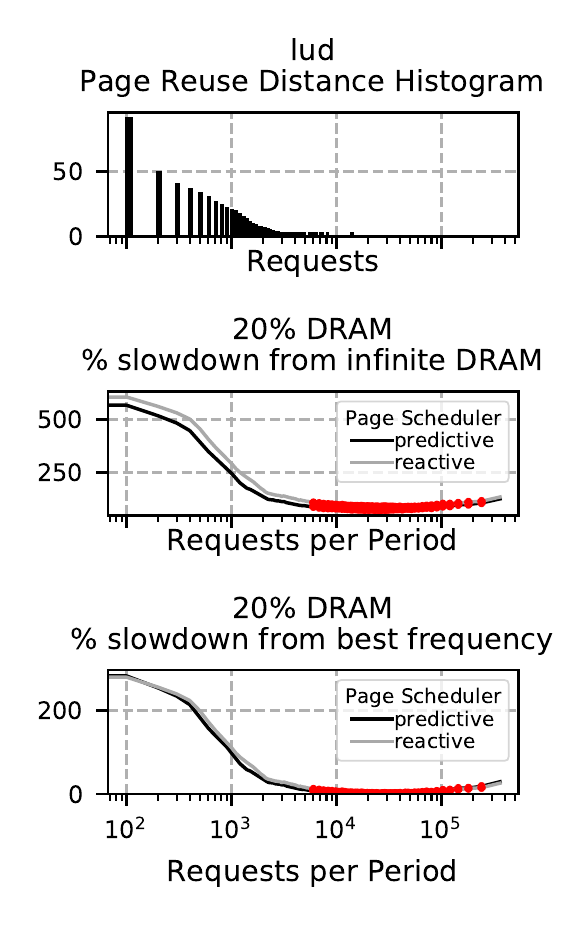}
\includegraphics[scale=0.6]{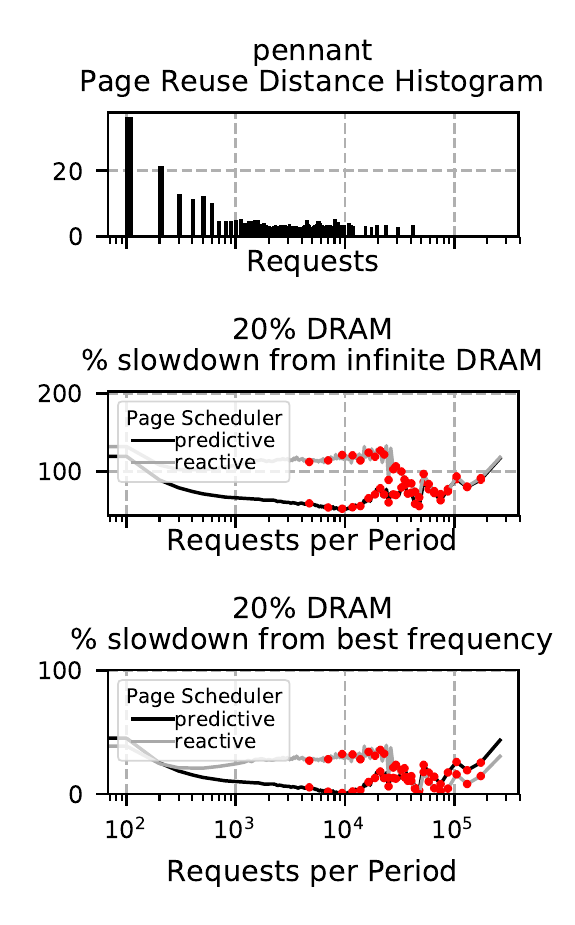}
\vspace{-2.0ex}
\caption{Histogram of page reuse distance and its relationship with application performance across period durations, for a predictive and reactive page scheduler over a simulated platform with DRAM and \nvm at a 20\%:80\% capacity ratio. The red dots correspond to the performance of the candidate frequencies generated by \sys.}
\vspace{-1.0ex}
\label{fig:reuse_perf}
\end{figure*}

\section{Motivation}
\label{sec:motivation}


\subsection{Performance Gap}
\label{sec:motiv-perf}

The big disparity in the proposed page scheduling frequencies across related works, summarized in Table \ref{tbl:related_work}, hints that they are empirically tuned to work best for their given page scheduling implementations and evaluated applications. For this reason, we capture the application performance gap created by using these proposed frequencies as opposed to an optimal frequency across a wide range of data access patterns. 
Figure~\ref{fig:perf_gap} captures application runtime slowdown from the case of an optimal frequency that provides best performance, together with the corresponding amount of data moved as a percentage of the application's memory footprint. The performance of our proposed solution \sys is also included in the figure, but will be further analyzed in Section \ref{sec:eval-benefit}. 

The proposed frequencies create a 10\%-100\% performance slowdown compared to the performance achievable with a best-case frequency, on average, across applications and page schedulers.
This makes a case for the need for a more robust tuning approach than the empirical one.
Taking a closer look, we observe that no single frequency works best across applications and page schedulers. In more detail, predictive vs. reactive page schedulers experience the lowest slowdown, on average, for frequencies that are an order of magnitude apart, that is a period duration of 1 second proposed by {\tt HMA} vs. 10 seconds by {\tt thermostat}, respectively. Additionally, the frequency that works best on average for a certain page scheduler may not provide best performance across {\it all} applications. For example, the lowest slowdown for a reactive page scheduler provided by {\tt thermostat} is not the best choice for {\tt pennant}, {\tt lud}, {\tt hotspot} and {\tt kmeans}. In particular, it incurs an average 8\% slowdown from the respective best {\it proposed} frequency, that is additional to the slowdown from the best frequency itself. 

\noindent{\bf Takeaways.} This initial experiment validates our concern that frequencies proposed by existing solutions leave a significant performance gap of 10\%-100\% across applications and page scheduler designs. No single proposed value works best across all applications and page schedulers. Therefore, there is an opportunity to close this performance gap with a more insightful tuning approach.

\subsection{Tuning Overheads}
\label{sec:motiv-complexity}

Existing solutions choose to empirically tune their page scheduling frequency and fix it across applications, to avoid the non-trivial tuning overheads of fine-grained frequency exploration. 
Stated more formally, 
an empirical tuning approach has $O(1)$ time complexity, since it chooses upon a constant set of frequencies. The choice of the frequencies themselves is critical, since an insight-less selection can lead to the aforementioned performance gap.

An exhaustive tuning approach has $O(N)$ time complexity, because the number of possible frequencies grows linearly with the application runtime. 
For example, the possible period durations for an application that generates $N$ memory requests in total, are the windows of any length between $[1, \frac{N}{2}]$, assuming that a page scheduler should run for at least two periods of $\frac{N}{2}$ requests each.
Similarly, if we consider the time domain instead of the memory request domain, the number of possible period durations is such that is splits the application runtime at multiples of a timestep, 
where a timestamp could be related to 
the Linux scheduling time slice, for instance.

\noindent{\bf The need for some insight.} The long runtime of applications that require massive hybrid memory systems makes an exhaustive tuning approach completely impractical. 
By using a more insightful tuning method we can drastically reduce these overheads, and also eliminate the performance gap caused by a poor choice of migration frequency made by empirical selection approaches.

\subsection{`Don't Break the Data Reuse' Insight}
\label{sec:motiv-reuse}

We perform the aforementioned exhaustive tuning approach to extract insights. We select applications with a wide range of data access behavior. Figure~\ref{fig:traces} shows a visual representation of their memory access patterns, as analyzed by the collected traces. We observe the strided array traversals of {\tt backprop} and {\tt quicksilver} vs. the distinctly shaped sparse tensor traversals of {\tt cpd}, the triangular multiplication in {\tt lud}, and the irregular memory accesses of {\tt pennant} over a fixed number of repetitive cycles.

\noindent{\bf Page Reuse Distance.} The top graphs in Figure~\ref{fig:reuse_perf} depict information on data reuse. In the context of
these analyses, we use page reuse distance as a measure for page reuse, where 
the page reuse distance is the number of memory accesses that are issued to other pages, between two consecutive accesses to a particular page. 
There is a clear connection between the page reuse distances and the access patterns in Figure~\ref{fig:traces}. For example, for {\tt backprop} the reuse distance of 20,000 requests maps to the gap between the large access strides, and it appears 15 times since there are 16 strides. 
In contrast, the decreasing appearances of page reuse distances for {\tt lud} and {\tt pennant} correspond to the triangular array traversal and random access behavior, respectively.

\noindent{\bf Relation of Performance and Data Reuse.} The bottom graphs in Figure~\ref{fig:reuse_perf} capture the application runtime slowdown from the case of infinite DRAM capacity and from the case of optimal frequency selection, across all possible period durations for predictive and reactive page schedulers. The x-axis is aligned with the histogram (top graph) and aims to capture the relation between the page reuse distances and page scheduling period durations.

We observe that predictive page schedulers, which make a better selection of which pages to move, provide best application performance for much shorter periods than reactive ones. However, irrespective of the page scheduler's effectiveness, very short periods create a significant aggregate data monitoring and movement overhead, as also shown in Figure~\ref{fig:perf_gap}. In addition, arbitrarily long periods do not allow the page scheduler to react promptly to changes in the access pattern behavior, thus create insufficient data movement to dynamically improve the data tiering. 
Moreover, the effectiveness of reactive page schedulers suffers at periods whose length is shorter than the page reuse distances with significant appearances, incurring an average of 50\% additional performance slowdown compared to predictive schedulers.
For example, this is the case for {\tt backprop} when periods are shorter than 20,000 requests per period, which is the page reuse distance of its strided access pattern. The scheduler's effectiveness drops because its reactive design identifies as hot pages the ones that correspond to a certain part of the access stride, then moves them to the limited DRAM capacity, but they will not be accessed in the next period, when the rest of the pages of the stride will be accessed. 
Such reactive page scheduling approaches are more effective when they operate over larger windows of access history, enabled either by longer periods or longer history of shorter periods. Regardless, the time window of access history should be large enough to not `break' the data reuse.

\noindent{\bf Lessons learned.} This extensive application performance 
characterization 
shows a clear relationship among the data reuse times and the page scheduling period durations which provide best performance. Reactive page schedulers benefit from periods that don't break the data reuse, to make better page migration decisions. Both reactive and predictive schedulers should avoid very short periods that reveal the data monitoring and movement costs, as well as arbitrarily long periods that do not allow a prompt response to changes in the data access pattern and create insufficient aggregate data movement. 

%% file: solution.tex
\begin{figure}
\centering
\includegraphics[scale=0.6]{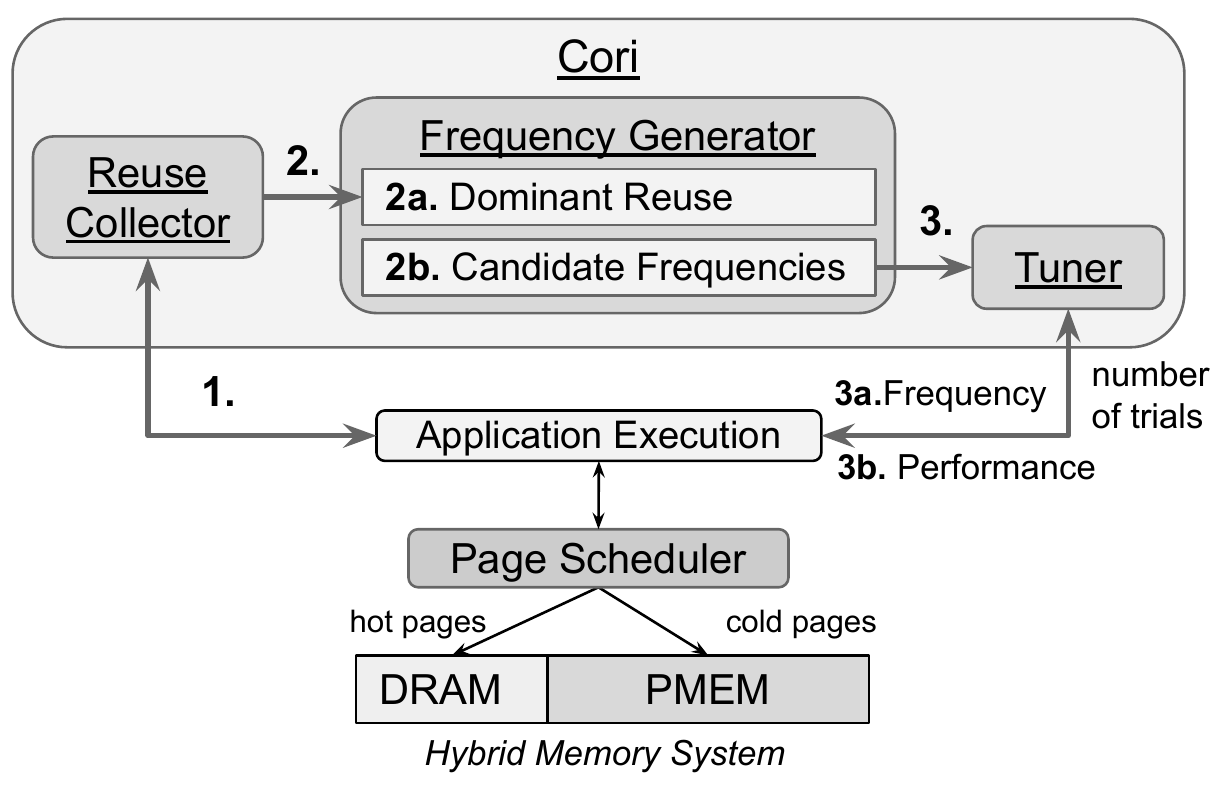}
\vspace{-1.0ex}
\caption{System components of \sys and its integration with the hybrid memory software stack.}
\vspace{-3.0ex}
\label{fig:system_design}
\end{figure}

\section{Solution}
\label{sec:solution}

\noindent{\bf Design Goals.} The objectives of our proposed frequency tuning solution are as follows:

\begin{tightitemize}
\item[\bf G1] {\it Bridge the performance gap} left by existing solutions that do not properly tune their page scheduling frequency.
\item[\bf G2] {\it Drastically reduce the number of tuning trials} needed to find the frequency that enables desired performance.
\item[\bf G3] {\it Build a generic tuning approach} that works across applications and page schedulers.
\item[\bf G4] {\it Enable practical system-level integration} using readily available information on application data access behavior, without explicit code-level modifications or specific APIs. 
\end{tightitemize}

\vspace{0.05in}
To address these goals, we propose {\bf \sys}, a method for tuning data movement frequency in hybrid memory systems.
\sys gleans data-movement requirements based on application-specific data reuse trends to guide the frequency tuning process, and select a frequency which delivers performance gains or increases in data movement efficiency ({\bf G1}) with a small number of tuning trials ({\bf G2}). 
\sys extracts the necessary information from execution
profiles,
and does not require any changes to applications or the memory management stack ({\bf G3}). 
Experimental results from a real testbed with DRAM and Intel Optane PMEM validate the simulation-bases evaluation of \sys, and demonstrate the feasibility of its system-level integration ({\bf G4}).

\vspace{0.05in}
\noindent{\bf \sys Overview.} Figure~\ref{fig:system_design} illustrates the system design of \sys and its interactions with the hybrid memory page scheduler, summarized as follows:
\begin{tightnumberenumerate}
\item The Reuse Collector executes a single profile run of the application to collect information on 
  data reuse.
\item The Frequency Generator analyzes the 
  data reuse profile and generates a range of proposed data movement frequencies.
  To achieve this, it first 
  calculates the dominant reuse period as a weighted average of the observed reuses (2a). 
  Then, it generates a range of candidate frequencies at time intervals that are multiples of the dominant reuse period (2b), and outputs the frequencies to the Tuner in decreasing order, from higher to lower frequencies, thus shorter to longer periods.
\item The Tuner makes a number of tuning trials with the candidate frequencies in the proposed order. 
  It configures the page scheduler to operate 
  at each of the recommended frequencies (3a). 
  It then observes the application runtime and resource use and determines whether the application performance has reached best or desired levels (3b). If not, the Tuner moves on to the next frequency in order, going back to step 3a.
\end{tightnumberenumerate}
\noindent Next, we describe in more detail these 
steps and system components.

\subsection{Reuse Collector}

The goal of the Reuse Collector component is to generate a histogram of data reuse similar to the ones 
shown in Section~\ref{sec:motiv-reuse}.
In the context of the simulation-based analysis we collect memory access traces and have access to detailed information on data reuse in terms of page reuse distances at the granularity of each individual memory access. This cannot be generally achieved for arbitrary applications, 
therefore, we propose a practical system-level alternative to collect similar information on data reuse.

\noindent{\bf Loop Durations.} We make the intuitive realization that data reuse appears mostly within loop operations during application execution. Therefore, information on the time duration of loops can be a practical estimation to page reuse distance in the time domain. Figure~\ref{fig:daniel-loops} depicts the time duration of loops across applications including {\tt backprop} and {\tt lud}. We observe a similar histogram shape to the ones generated via the memory access traces for the page reuse distances in Figure~\ref{fig:reuse_perf}:
{\tt backprop} has distinct loop durations that repeat around 15 times, which corresponds to the 16 data access strides depicted in Figure~\ref{fig:traces},
and {\tt lud} shows a gradual degradation in the loop durations due to the triangular array traversal and decreasing reuse of the number of pages shown in the same figure.
We validate that the loop duration histograms of
the remaining applications match what we 
observed via the memory access trace collection.

\noindent{\bf Collection of Loop Durations.} In the context of validating \sys on a native testbed in Section~\ref{sec:eval-validate}, we instrumented the applications source code and individually timed the duration of the primary for loops. In principle, however, such instrumentation can easily be performed using compiler-level~\cite{polly,beacons:pact15} or binary instrumentation techniques~\cite{loopprof,dyninst}. In the current paper, we do not present a complete \sys tool which integrates such techniques, rather we focus on establishing the methodology that forms the basis of such a tool, and demonstrate via manual instrumentation that the methodology is effective.
 We verify that we can obtain accurate loop timings using a LLVM compiler pass, similar to what has been used as part of the Beacons compiler framework~\cite{beacons:pact15}, which automatically generates the instrumented binary without any application source code modifications.

\subsection{Frequency Generator}

\noindent{\bf Dominant Reuse.} The Frequency Generator analyzes the data reuse histogram provided by the Reuse Collector, in order to identify the one that best represents the range of captured reuses. We refer to this as the {\it dominant reuse}.
Dominant reuse (DR) is computed as a weighted average of the observed data reuses ($N$ different reuses) in the histogram, as summarized in Equation~\ref{eq:dom-reuse}. The weights are the number of appearances $repeat_i$ of a reuse $reuse_i$ in the corresponding histogram. This will shift the average towards the data reuse distances that repeat more times. Additionally, we introduce an extra weight $(N-i)$ that favors shorter reuse distances, because this will allow us to generate a more calibrated selection of candidate frequencies, that works irrespective of the page scheduler's effectiveness,
as we evaluate in Section~\ref{sec:evaluation}. 

\begin{equation} \label{eq:dom-reuse}
\centering
DR =  \frac{\sum_{i=1}^{N} (N-i) \times repeat_{i} \times reuse_{i}}{\sum_{i=1}^{N} (N-i) \times repeat_{i}}
\end{equation}
\begin{equation} \label{eq:candidate-periods}
CandidatePeriods =  [DR, \: 2 \times DR, \: 3 \times DR, \: ..., \: \frac{Runtime}{2}]
\end{equation}

\noindent{\bf Output Candidate Frequencies.}
Based on DR, 
the Frequency Generator creates a sequence of candidate data movement periods at time intervals that are multiples of DR, 
as shown in Equation~\ref{eq:candidate-periods}. The last possible candidate in the sequence is the one that splits in half the overall application runtime that the Reuse Collector has previously observed. 
The candidate frequencies are derived by simply inverting the values of the candidate periods.
Figure~\ref{fig:reuse_perf} includes a visual representation of the candidate periods as red dots. 
Finally, the Frequency Generator outputs to the Tuner the candidate frequencies in the specified order from shorter to longer periods, thus higher to lower data movement frequencies. This priority ordering, together with the dominant reuse calculation, is essential to \sys's success, compared to other possible solutions, as we evaluate upon in Section~\ref{sec:eval-cost}.

\subsection{Tuner}
The Tuner uses the sequence of candidate frequencies to perform the actual tuning procedure. The Tuner starts its initial trial with the first frequency in order, sets it as the operational page scheduling frequency and executes the application over the hybrid memory. If performance is within desired levels or the best one observed (after the first trial), the Tuner chooses to stop or continue the tuning process. When the Tuner finds the frequency that provides best performance after a number of trials, the selected frequency is kept for any subsequent execution of the particular application on the given combination of platform configuration and page scheduler.

\subsection{Discussion}

\sys currently improves upon tuning approaches, such as the empirical ones,
by observing best performance across a number of tuning trials of actual application execution. The decision of after how many trials the tuning stops is flexible. There can be a fixed number of trials or tuning can stop after performance reaches desired levels or shows no significant variation from the last trial. However, such an execution-based tuning methodology may be impractical for long running applications, such as training machine learning models and scientific simulations. Nonetheless, \sys only requires the collection of data reuse information, that can be made readily available using compiler-assisted instrumention, 
laying the grounds towards an online frequency tuning solution. \sys can be extended with system-level performance metrics and combined with online access pattern detection solutions used in prefetching \cite{LMA, leap}, or machine intelligent page schedulers \cite{kleio}, so as to adapt the page migration frequency to dynamic changes in data reuse and access patterns.
Finally, the recommendations made by \sys depend on the calculation of the dominant reuse, and are therefore sensitive to the granularity at which the data reuse information is collected and aggregated. The evaluations presented in this paper base the calculation on reuse information captured at granularity of 1000s of data accesses (in the simulation framework) and of each loop (on the real hardware testbed). This instrumentation granularity can be dynamically adjusted to trade among the tuning overheads vs.~the quality of the recommendations.

%% file: evaluation.tex
\section{Evaluation}
\label{sec:evaluation}

The goal of the evaluation is to demonstrate how \sys realizes its design goals. First, we highlight the benefit of using Cori with respect to application performance improvements and system resource efficiency. Second, we evaluate  the tuning overheads of using \sys. Finally, we validate the effectiveness and practicality of \sys on the native Intel Optane platform.

\subsection{Benefit of Using \sys}
\label{sec:eval-benefit}

Figure~\ref{fig:perf_gap} includes the application performance and data moved when the page scheduling frequency is tuned with \sys, compared to the frequencies proposed by existing solutions. Regarding application performance, using the frequencies selected by \sys achieves on average a 3\% slowdown compared to when using the best possible frequency for each of the applications. In comparison, the frequencies used by other techniques result in an average 10\%-100\% slowdown
from the ideal case. For cases where \sys does not provide the best application performance, as in the case of {\tt quicksilver} with a predictive page scheduler, 
the performance with \sys is less than 3\% away from the best observed one.
As discussed in Section~\ref{sec:motiv-perf}, no other set of frequencies proposed by existing solutions 
provides as good performance across all applications {\it and} page schedulers, as \sys. 

In this experiment, the frequency tuning in \sys is performed to optimize application performance, so it is not surprising that it does not provide uniformly lowest data movement. However, 
\sys realizes the necessary data movements to achieve the provided application performance levels. For instance, the increased data movement compared to some of the predictive schedulers (e.g., 3$\times$ more compared to thermostat), are offset by the reductions in the slowdown compared to the best frequency case (5$\times$). 
We highlight the actual number of GBs moved for a native hybrid memory platform later in Section~\ref{sec:eval-validate}. 

{\it \sys meets the {\bf G1} design goal by bridging the performance gap left by existing solutions and achieves only a 3\% average slowdown from an optimal frequency selection across applications and page schedulers.}

\begin{figure}
\begin{subfigure}{\linewidth}
  \centering
  \includegraphics[scale=0.55]{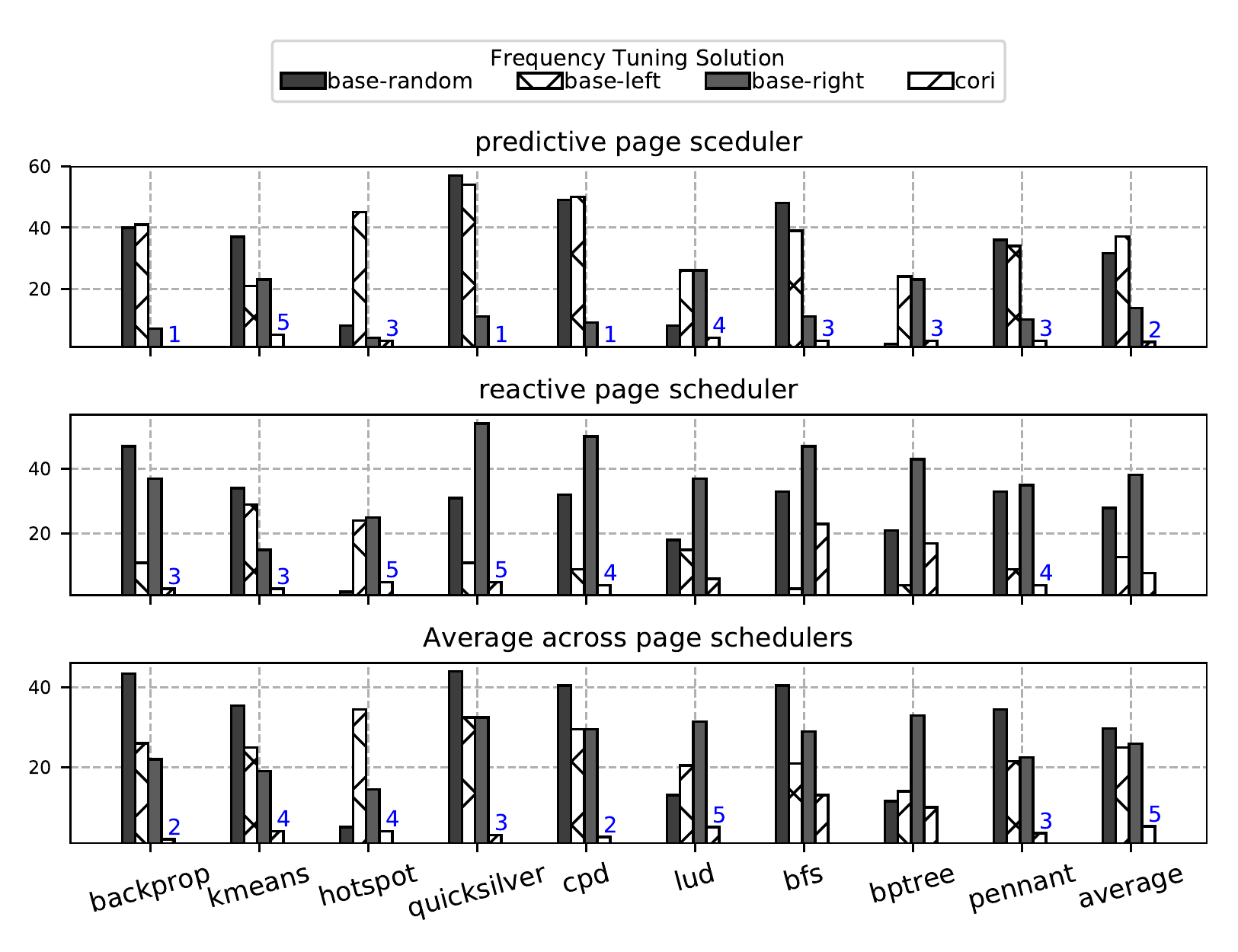}  
  \vspace{-2.0ex}
  \caption{Number of tuning trials to find best performance. \sys (blue text) requires the minimum number of trials on average across all applications {\it and} page schedulers.}
  \label{fig:eval_compare_num_trials}
\end{subfigure}

\begin{subfigure}{\linewidth}
  \centering
\includegraphics[scale=0.55]{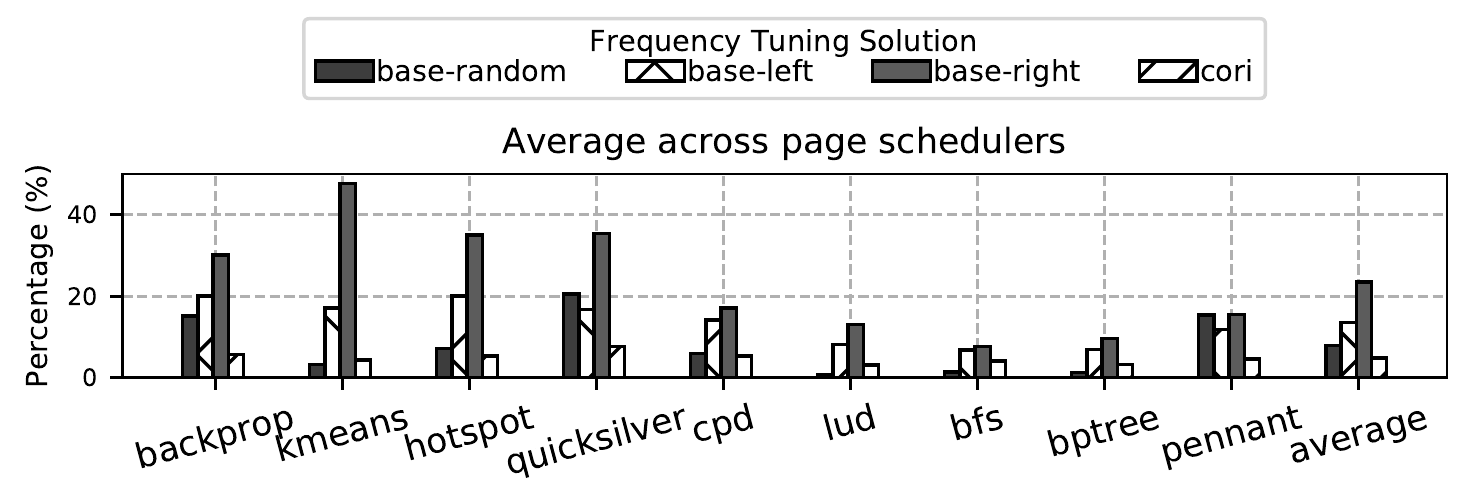}
  \vspace{-1.0ex}
  \caption{Performance slowdown from best frequency for Cori's number of trials. \sys is the only solution that provides lowest slowdown consistently for {\it every} application {\it and} page scheduler.}
  \label{fig:eval_compare_perf}
\end{subfigure}

\begin{subfigure}{\linewidth}
  \centering
\includegraphics[scale=0.55]{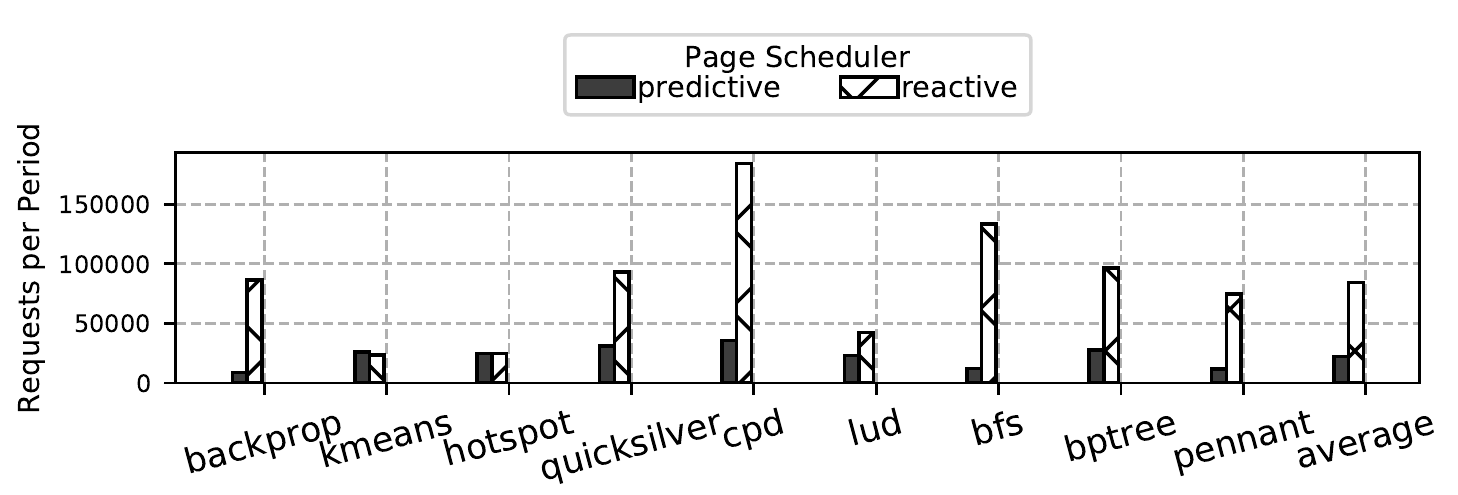}
  \vspace{-1.0ex}
  \caption{Final period time duration selection of \sys.}
  \label{fig:eval_freqs}
\end{subfigure}

\caption{Comparison of \sys with a baseline frequency tuning solution for a simulated hybrid memory system with DRAM and PMEM at 20\%:80\% capacity ratio.}
  \vspace{-3.0ex}
\label{fig:eval_compare}
\end{figure}

\subsection{Overhead of Using \sys}
\label{sec:eval-cost}
We evaluate the overheads of using \sys by comparing the number of tuning trials required by \sys to find the best frequency vs.~what is required to find that value using other tuning methods. We also evaluate whether \sys's overheads are justified, by comparing how close \sys is to the performance of a system which operates at the best possible frequency for each of the applications, vs.~how close would the other methods be if they use the same number of trials as \sys. 

Given the lack of a non-empirical tuning approach, we construct a {\bf baseline}, which like \sys, operates at the system-level, but is blind to any insights it might have regarding application requirements. 
This baseline explores the $O(N)$ problem space of all possible frequencies
by using a simple step function, 
with 
candidate periodic time intervals that differ by a duration of $timestep$, as summarized in Equation~\ref{eq:baseline-freqs}, 
The corresponding frequencies are derived by inverting the periodic time intervals.
\begin{equation} \label{eq:baseline-freqs}
Base \: Candidates =  [timestep, \: 2 \times timestep, \: ..., \: \frac{Runtime}{2}]
\end{equation}

Next, 
we vary the {\bf priority ordering} of the generated candidate frequencies. First, the {\tt base-left} baseline starts from low frequencies (large periods) and moves to the left towards higher frequencies (short periods) in the sequence described in Equation~\ref{eq:baseline-freqs}.
The {\tt base-right} baseline starts from high frequencies and moves towards the right to lower ones, similar to \sys. Third, we also assume a {\tt base-random} approach that randomly explores values in the sequence. 


Figure~\ref{fig:eval_compare_num_trials} shows the number of tuning trials required for best application performance. Among the baseline variations none of them works best across all applications {\it and} page schedulers. More specifically, while {\tt base-right} is the baseline approach that requires least trials for a predictive page scheduler across application,  {\tt base-left} works best for reactive page schedulers.
Thus, a baseline approach that explores frequencies in a certain priority ordering needs further insights to identify best performance in a reasonable number of tuning trials. Even though  {\tt base-random} is independent of such a priority ordering, its unpredictable frequency selection results in worst-case average tuning overheads. 

In contrast, the guided frequency selection performed by \sys, allows it to make a recommendation in just two trials
on average for predictive page schedulers. Across applications and page schedulers \sys reduces the number of trials by {\bf 5$\times$},
from 25 on average across baselines down to only 5 trials. The only corner case where \sys requires up to 20 trials is for applications with random access patterns like {\tt bfs} and {\tt bptree} where the access pattern prediction capabilities of a reactive page scheduler are limited. This is the reason why for such applications, when a more predictive page scheduler \cite{kleio} is not available, the cache organization of the hybrid memory is shown to be more beneficial~\cite{optane-vldb}.

{\it \sys meets the {\bf G2} design goal by reducing by 5$\times$ the number of tuning trials needed to reach an average of only 3\% performance slowdown, compared to  baselines that ignore insights about application requirements,  and as low as 2 trials on average for predictive page schedulers.}


Figure~\ref{fig:eval_compare_perf} shows the performance that the baselines provide
when executing for the same number of tuning trials that \sys requires to find best performance. The values are averaged 
across the page schedulers. On average, the baselines incur higher performance slowdown because they require significantly more trials to reach best performance, as shown in Figure~\ref{fig:eval_compare_num_trials}. 
Within the execution overhead of \sys, only the {\tt base-random} approach seems to be able to still choose frequencies that provide good performance, but only for some of the applications; for others (e.g., {\tt quicksilver} and {\tt pennant}), {\tt base-random} is less effective even compared to some of the other baselines.

For completeness, in Figure~\ref{fig:eval_freqs}, we also show the best frequencies selected by \sys for the two types of schedulers.
We highlight that the best frequency that maximizes application performance differs across schedulers and across applications, further justifying the use of \sys. 

{\it \sys meets the {\bf G3} design goal since it 
  provides maximum performance improvements for minimum number of tuning trials across applications and page schedulers.}

\begin{figure}
\begin{subfigure}{\linewidth}
  \centering
  \includegraphics[scale=0.55]{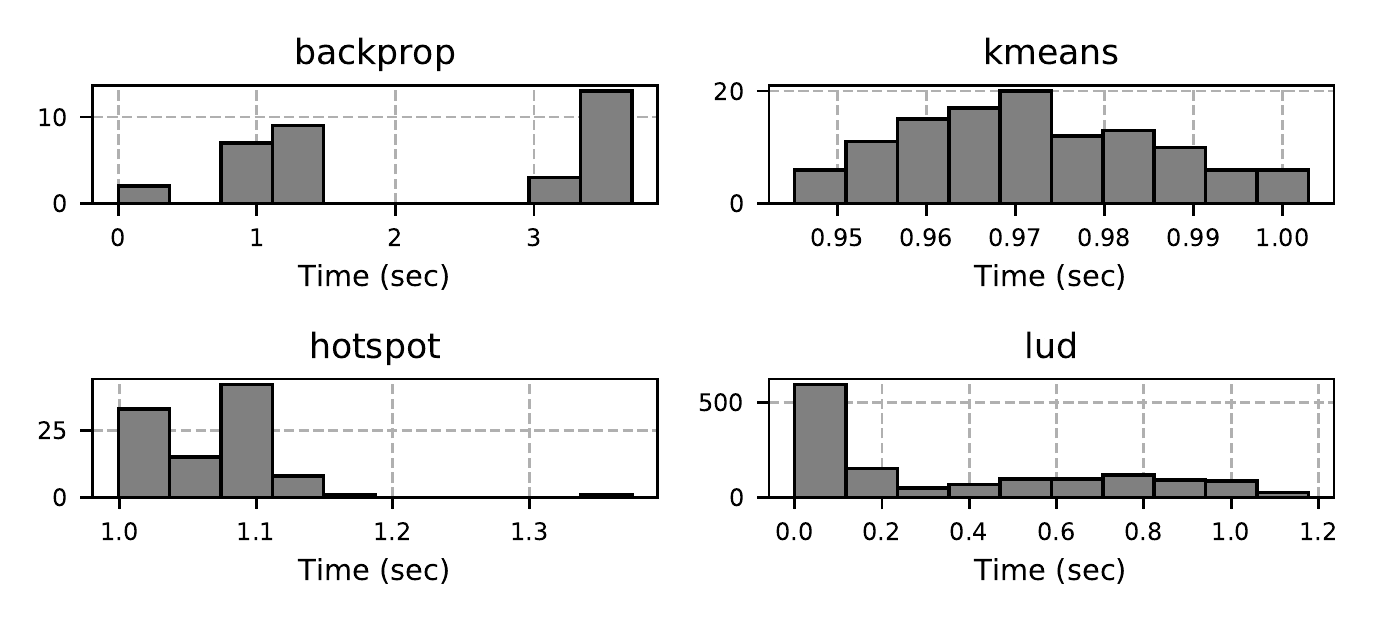}  
  \vspace{-2.0ex}
  \caption{{\bf \sys Step 1:} Collect loop time durations.}
  \label{fig:daniel-loops}
\end{subfigure}

\begin{subfigure}{.5\linewidth}
  \centering
  \includegraphics[scale=0.5]{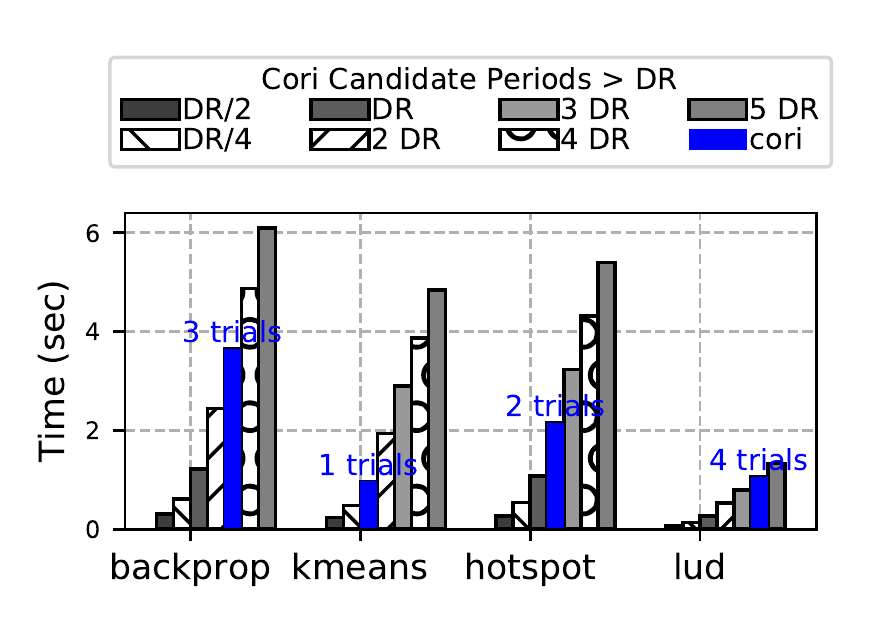}
  \caption{{\bf \sys Step 2:} Calculate the dominant reuse DR and generate candidate period lengths at multiples of the DR. {\bf Final Choice:} Select the first period length (blue bar) after x trials, that brings lowest runtime {\it and} migrations.}
  \label{fig:daniel-candidates}
\end{subfigure}~
\begin{subfigure}{.5\linewidth}
  \centering
\includegraphics[scale=0.5]{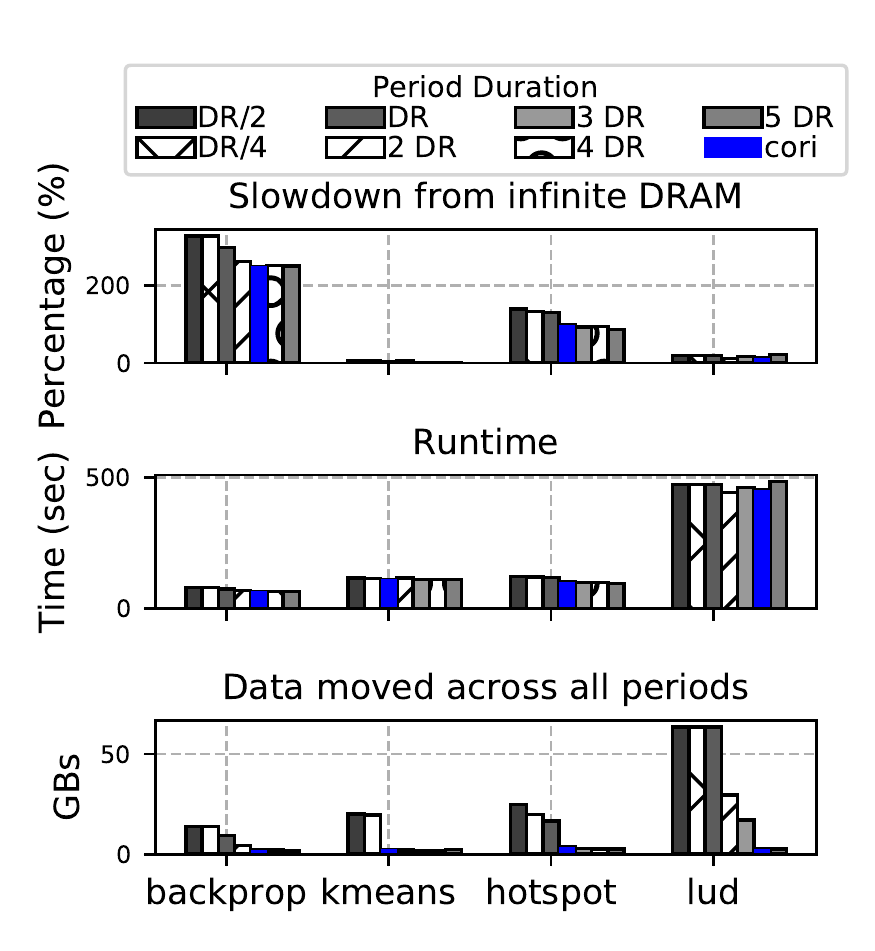}
  \caption{{\bf \sys Step 3:} Tuning trials of application performance.}
  \label{fig:daniel-perf}
\end{subfigure}
\caption{System-level validation of \sys for a reactive page scheduler that executes on a native Optane DC PMEM platform for 20\%:80\% DRAM:PMEM capacity ratio.}
\label{fig:daniel-data}
\vspace{-2.0ex}
\end{figure}

\subsection{Optane DC PMEM Validation}
\label{sec:eval-validate}

We validate the simulation-based observations about \sys by providing results from a native hybrid memory platform. These experiments also demonstrate the feasibility of using \sys as a practical system-level solution for frequency tuning. The experiments are conduced on an Intel Optane platform, configured with 20\%:80\% DRAM to PMEM capacity ratio, and 
a reactive page scheduler kernel module that operates over a window of past access history, both as described in Section~\ref{sec:method-optane}. Then, we go through the steps of \sys as summarized in Figure~\ref{fig:system_design} and report our findings in Figure~\ref{fig:daniel-data}.

\noindent{\bf Recreating \sys's steps.}
First, we gather information on data reuse. More specifically, we collect the time duration of the loops across applications, as shown in Figure~\ref{fig:daniel-loops}, using the suggested approach in Section~\ref{sec:solution}. 
Second, we calculate the dominant reuse as described in Equation~\ref{eq:dom-reuse} and generate the candidate period durations at multiples of the dominant reuse, as shown in Figure~\ref{fig:daniel-candidates}. While for {\tt backprop, kmeans, hotspot} the dominant reuse is around 1 sec, for {\tt lud} it is much less, given the corresponding loop duration histogram.  We also include period durations that are less than the dominant reuse, to validate whether performance indeed is not best for such periods that \sys does not include in its sequence of candidates.
Third, we replicate \sys's tuning process by executing the applications for the selected period durations in increasing order and observe the runtime, its slowdown from the ideal case of infinite DRAM and data moved, as shown in Figure~\ref{fig:daniel-perf}. The final choice according to \sys is the {\it first} period duration in the experimentation order that drastically reduces the amount of data moved and thus appropriately reduces the runtime. Figure \ref{fig:eval_freqs} inidicates in blue the final period choice and the number of tuning trials it required. 

\noindent{\bf Validation observations.} First, we observe that period lengths that are shorter than the dominant reuse (DR/4, DR/2), create tens of GBs of more data moved, consistently across all applications. 
This confirms
the insight presented in Section~\ref{sec:motiv-reuse} that the operational period should not be shorter than the data reuse pattern. 
Also, it validates \sys's effectiveness in calculating the dominant reuse and choosing it as the initial point of tuning. The performance with much larger periods is not included in Figure~\ref{fig:daniel-perf}, since it can be substantially worse, such as 50\% of runtime slowdown for {\tt lud} at 5 second periods, and \sys's tuning ends at much shorter periods.

Second, regarding application performance and system resource efficiency, \sys selects the period duration that reduces to their lowest levels both the data moved and the runtime slowdown from the case of infinite DRAM capacity, across all applications.
For some applications these levels of runtime slowdown are less significant than others.
For applications like {\tt kmeans} and {\tt lud} very short periods that force the reactive page scheduler to create a burst of asynchronous data movements, are not enough to stress the Optane's bandwidth and proportionally reflect on their runtime. Regardless, \sys identifies the page scheduling frequency that enables the best performance levels allowed by the available DRAM capacity and
minimizes data movement overheads. 
Additionally, the levels of runtime slowdown observed in this experiment, are very similar to the ones captured in our simulation (Figure~\ref{fig:reuse_perf}), validating its correctness.

Finally, the selected periods themselves are different across applications and range between 1-3 seconds. Even though this doesn't seem as a substantial difference, 
empirical approaches may have ignored values in such proximity,
however, 
for {\tt backprop} the runtime slowdown reduces by 50\% when going from 1 second to 3 second periods, and for {\tt hotspot} by 30\% when switching from 1 second to 2 second periods. This validates the benefit from using \sys toward realizing significant application performance improvements, within only 2-3 average tuning trials, minimizing the tuning overheads. 

{\it \sys meets the {\bf G4} design goal by allowing for a practical integration with existing hybrid memory system-level managers, and can be realized without modification to applications and system-level components. Validation of \sys on the Intel Optane DC PMEM platform, confirms the simulation-based motivational arguments and insights, and highlights the benefit of using \sys in return for minimal tuning overhead.}

%% file: related.tex
\section{Related Work}
\label{sec:related}

\noindent{\bf Data Tiering across Hybrid Memory.} There is a wide range of data tiering solutions for hybrid memory systems configured as a flat memory address space, that operate across levels of the memory management stack. First, at the application-level there are solutions that propose custom data allocation APIs to improve initial and dynamic data placement \cite{xmem, memkind}. Second, at the runtime-level solutions instrument MPI communication phases and task-based parallel execution to 
initiate 
and synchronize the data movement \cite{Unimem, tahoe}. Moving to the operating system-level (or hypervisor-level) there are solutions which perform periodic data movements using the current NUMA-based page migration support \cite{thermostat, nimble, hetero-os, hetero-visor, nimble}, or appropriately extend NUMA-based data balancing policies \cite{hinuma}. Finally,  hardware-assisted solutions aim to reduce the data monitoring and movement software overheads \cite{utility, HMA, mempod}. Our work targets operating system-level periodic data tiering solutions and tunes their operational frequency.

\noindent{\bf Tuning of System Parameters.}
The opportunities for improving performance and efficiency via careful tuning of system-level parameters have been established and demonstrated across different contexts, 
ranging from operating system-level configuration parameters \cite{linux-server-tuning}, voltage-frequency balancing for power management \cite{dvfs}, CPU scheduling \cite{Seltzer2006OperatingSS}, and database index tuning \cite{cophy}. Such traditional observation-driven tuning techniques are being replaced by reinforcement learning \cite{capes, qtune},
and more generally, learning-augmented solutions
are being developed across the systems software stack, and  
can be unified using machine intelligence frameworks such as AutoSys \cite{autosys}. 
\sys 
uses a traditional observation-driven tuning approach and targets optimizations of the hybrid memory page scheduling frequency, 
however, the insight from this work can be incorporated in future machine learning-based approaches. 

%% file: summary.tex
\section{Conclusion}
\label{sec:summary}

This work presents \sys, a system-level solution for tuning the operational frequency of data tiering solutions that periodically move data across flat hybrid memory components. \sys synthesizes insights on data reuse information to better guide the process of selecting frequency candidates, 
reducing by, on average, 5$\times$ the number of tuning trials from an insight-less exploration. This way, \sys delivers performance improvements within 3\% from the case of optimally chosen frequency, completely eliminating the 10\%-100\% performance gap created by using operational frequencies adopted across recent related works. \sys is robust, and provides benefits across application data access patterns and page migration policies. Importantly, we validate that \sys's approach is effective in the context of a real system, and that it can be integrated as part of a system-level tuning solution, via the use of information that can be automatically extrated through compiler-based methods, without requiring any code-level modifications of applications or across the memory management software stack; our next steps will enhance the system with such capabilities.